\documentclass[11pt,twoside]{article}
\usepackage{CAGN2016}
\usepackage{graphicx}

\usepackage[T1]{fontenc} 

\usepackage{latexsym}
\usepackage{verbatim}

\usepackage{hyperref}
\usepackage{xspace}
\usepackage[usenames, dvipsnames]{xcolor}

\newcommand{\etal}{et\thinspace al.\thinspace}
\newcommand{\hii}{H\thinspace\textsc{ii}\xspace}
\newcommand{\bond}{\textsc{bond}\xspace}

\newcommand{\Ha}{\ifmmode \mathrm{H}\alpha \else H$\alpha$\fi\xspace}
\newcommand{\Hb}{\ifmmode \mathrm{H}\beta \else H$\beta$\fi\xspace}
\newcommand{\neiii}{\ifmmode [\mathrm{Ne}\,\textsc{iii}] \else [Ne~{\scshape iii}]\fi\xspace}
\newcommand{\Neiii}{\ifmmode [\mathrm{Ne}\,\textsc{iii}]\lambda 3869 \else [Ne~{\scshape iii}]$\lambda 3869$\fi\xspace}
\newcommand{\oii}{\ifmmode [\mathrm{O}\,\textsc{ii}] \else [O~{\scshape ii}]\fi\xspace}
\newcommand{\Oii}{\ifmmode [\mathrm{O}\,\textsc{ii}]\lambda 3726 + \lambda 3729 \else [O~{\scshape ii}]$\lambda 3726 + \lambda 3729$\fi\xspace}
\newcommand{\Oiiit}{\ifmmode [\mathrm{O}\,\textsc{iii}]\lambda 4363 \else [O~{\scshape iii}]$\lambda 4363$\fi\xspace}
\newcommand{\heii}{\ifmmode \mathrm{He}\,\textsc{ii} \else He~{\scshape ii}\fi\xspace}
\newcommand{\Heii}{\ifmmode \mathrm{He}\,\textsc{ii}\lambda 4686 \else He~{\scshape ii}$\lambda 4686$\fi\xspace}
\newcommand{\ariv}{\ifmmode [\mathrm{Ar}\,\textsc{iv}] \else [Ar~{\scshape iv}]\fi\xspace}
\newcommand{\Ariv}{\ifmmode [\mathrm{Ar}\,\textsc{iv}]\lambda 4711 + \lambda 4740 \else [Ar~{\scshape iv}]$\lambda 4711 + \lambda 4740$\fi\xspace}
\newcommand{\Niit}{\ifmmode [\mathrm{N}\,\textsc{ii}]\lambda 5755 \else [N~{\scshape ii}]$\lambda 5755$\fi\xspace}
\newcommand{\hei}{\ifmmode \mathrm{He}\,\textsc{i} \else He~{\scshape i}\fi\xspace}
\newcommand{\Hei}{\ifmmode \mathrm{He}\,\textsc{i}\lambda 5876 \else He~{\scshape i}$\lambda 5876$\fi\xspace}
\newcommand{\nii}{\ifmmode [\mathrm{N}\,\textsc{ii}] \else [N~{\scshape ii}]\fi\xspace}
\newcommand{\niis}{\ifmmode [\mathrm{N}\,\textsc{ii}]_\mathrm{S} \else [N~{\scshape ii}]$_\mathrm{S}$\fi\xspace}
\newcommand{\Nii}{\ifmmode [\mathrm{N}\,\textsc{ii}]\lambda 6584 \else [N~{\scshape ii}]$\lambda 6584$\fi\xspace}
\newcommand{\oiii}{\ifmmode [\mathrm{O}\,\textsc{iii}] \else [O~{\scshape iii}]\fi\xspace}
\newcommand{\oiiis}{\ifmmode [\mathrm{O}\,\textsc{iii}]_\mathrm{S} \else [O~{\scshape iii}]$_\mathrm{S}$\fi\xspace}
\newcommand{\Oiii}{\ifmmode [\mathrm{O}\,\textsc{iii}]\lambda 5007 \else [O~{\scshape iii}]$\lambda 5007$\fi\xspace}
\newcommand{\sii}{\ifmmode [\mathrm{S}\,\textsc{ii}] \else [S~{\scshape ii}]\fi\xspace}
\newcommand{\Sii}{\ifmmode [\mathrm{S}\,\textsc{ii}]\lambda 6716 + \lambda 6731 \else [S~{\scshape ii}]$\lambda 6716 + \lambda 6731$\fi\xspace}
\newcommand{\ariii}{\ifmmode [\mathrm{Ar}\,\textsc{iii}] \else [Ar~{\scshape iii}]\fi\xspace}
\newcommand{\Ariii}{\ifmmode [\mathrm{Ar}\,\textsc{iii}]\lambda 7135 \else [Ar~{\scshape iii}]$\lambda 7135$\fi\xspace}
\newcommand{\siii}{\ifmmode [\mathrm{S}\,\textsc{iii}] \else [S~{\scshape iii}]\fi\xspace}
\newcommand{\Siii}{\ifmmode [\mathrm{S}\,\textsc{iii}]\lambda 9069 \else [S~{\scshape iii}]$\lambda 9069$\fi\xspace}

\newcommand{\rOii}{ \ifmmode [\mathrm{O}\,\textsc{ii} ]\lambda 3726/3729 \else [O~{\scshape  ii}]$\lambda 3726/3729$\fi\xspace}
\newcommand{\rOiii}{\ifmmode [\mathrm{O}\,\textsc{iii}]\lambda 4363/5007 \else [O~{\scshape iii}]$\lambda 4363/5007$\fi\xspace}
\newcommand{\rAriv}{\ifmmode [\mathrm{Ar}\,\textsc{iv}]\lambda 4740/4711 \else [Ar~{\scshape iv}]$\lambda 4740/4711$\fi\xspace}
\newcommand{\rNii}{ \ifmmode [\mathrm{N}\,\textsc{ii} ]\lambda 5755/6584 \else [N~{\scshape  ii}]$\lambda 5755/6584$\fi\xspace}
\newcommand{\rSiii}{\ifmmode [\mathrm{S}\,\textsc{iii}]\lambda 6312/9532 \else [S~{\scshape iii}]$\lambda 6312/9532$\fi\xspace}
\newcommand{\rSii}{ \ifmmode [\mathrm{S}\,\textsc{ii} ]\lambda 6731/6717 \else [S~{\scshape  ii}]$\lambda 6731/6717$\fi\xspace}

\newcommand{\qh}{$Q({\mathrm{H^{0}}})$\xspace}
\newcommand{\qhe}{$Q({\mathrm{He^{0}}})$\xspace}

\newcommand{\Nepp}{Ne$^{++}$}
\newcommand{\Arpp}{Ar$^{++}$}
\newcommand{\Opp}{O$^{++}$}
\newcommand{\Np}{N$^{+}$}

\begin{document}

\vskip 1.0cm
\markboth{N.~Vale Asari et al.}{A quick start guide to BOND}
\pagestyle{myheadings}
%
%
\vspace*{0.5cm}
\parindent 0pt{Contributed  Paper}


\vspace*{0.5cm} \title{A quick-start guide to {\scshape bond}: Bayesian
  Oxygen and Nitrogen abundance Determinations in \hii regions using
  strong and semistrong lines}

\author{N.~Vale Asari$^{1}$,
         G.~Stasi\'nska$^{2}$,
         C.~Morisset$^{3}$, and
         R.~Cid Fernandes$^{1}$}
\affil{$^{1}$Departamento de F\'{\i}sica--CFM, Universidade Federal de
       Santa Catarina, C.P.\ 476, 88040-900, Florian\'opolis, SC, Brazil \\
       $^{2}$LUTH, Observatoire de Paris, CNRS, Universit\'e Paris
       Diderot; Place Jules Janssen 92190 Meudon, France \\
       $^{3}$Instituto de Astronom{\'\i}a, Universidad Nacional
       Aut\'onoma de M\'exico, Apdo. Postal 70264, M\'exico D.F., 04510 M\'exico \\
}

\begin{abstract}
  We present a quick-start guide to \bond, a statistical method to
  derive oxygen and nitrogen abundances in \hii regions.  \bond compares
  a set of carefully selected strong and semistrong emission lines to a
  grid photoionization models.  The first novelty, in comparison to
  other statistical methods, is that \bond relies on the \ariii/\neiii
  emission line ratio to break the oxygen abundance bimodality. In doing
  so, we can measure oxygen and nitrogen abundances without assuming any
  a priori relation between N/O and O/H.  The second novelty is that
  \bond takes into account changes in the hardness of the ionizing
  radiation field, which can come about due to the ageing of \hii
  regions or the stochastically sampling of the IMF.  We use the
  emission line ratio \hei/\Hb, in addition to commonly used strong
  lines, to constrain the hardness of the ionizing radiation
  field. Finally, we also stress the pragmatic considerations behind our
  Bayesian inference.
\end{abstract}

\section{Why a statistical method based on photoionization models}
\label{ojo}

When direct temperature measurements are missing, statistical methods
are used to infer abundances in giant \hii regions. There are two
families of statistical methods. One is based on calibrating samples of
objects for which the abundance could be derived from temperature-based
methods. The other is based on photoionization model grids. The latter
is free from observational biases, but the grid must cover all the
configurations that could be found in nature. \bond (Bayesian Oxygen and
Nitrogen abundance Determinations) belongs to the second family.  \bond
infers oxygen and nitrogen abundances using carefully selected strong
and semistrong lines by comparing them to a grid of photoionization
models.  The source code is open and freely available at
\url{http://bond.ufsc.br/}. Full details can be found at Vale Asari
\etal (2016).  This manuscript is intended to be a quick-start guide to
highlight the most important aspects of the method.

\section{What sets \bond apart}

Common strong line methods based on simple calibrations (\oiii/\nii,
\nii/\Ha, $(\oii+\oiii)/\Hb$) assume that emission-line nebulae are a
one-parameter family, and that such parameter is the oxygen abundance.
The first to realise the importance of introducing a secondary parameter
to measure abundances was McGaugh (1991), who considered the effect of
the ionization parameter $U$.  Nowadays there is a plethora of methods
to measure abundances based on the comparison of observed to theoretical
emission lines from a grid of photoionization models (McGaugh 1991;
Kewley \& Dopita 2002; Tremonti \etal 2004; Dopita \etal 2013;
P\'erez-Montero 2014; Blanc \etal 2015).  We will discuss the difference
of our method with respect to others.

One novelty in our grid of photoionization models is that we do not
impose any a priori relation between N/O and O/H. The only other method
that also does not tie in the nitrogen and oxygen abundances is the one
by P\'erez-Montero (2014), with the difference that his method uses
auroral lines (so it is not a strong line method), and it selects models
around the empirical N/O versus O/H relation if auroral lines are not
available.

Another novelty in our method is that it uses Bayesian inference to
measure abundances, not quite unlike the one by Blanc \etal (2015). The
curious reader is referred to Section \ref{sec:bayes} for the key
points; we warn that, although Bayesian inference is part and parcel of
our method, it is not the most important aspect of \bond.

The killer features in \bond are: (a) N/O is free to vary, (b) it
distinguishes between the lower and upper metallicity branches, and (c)
it considers the effect of varying the hardness of the ionizing
radiation field. The next section shows how we have tackled (a), (b) and
(c), and explains the reasoning behind choosing which emission lines
need to be fitted.

\section{Input emission lines for \bond}

Our grid of photoionization models spans a wide range in O/H, N/O, and
$U$. The ionizing radiation field is provided by the instantaneous
starburst models of Moll\'a \etal (2009) for six different ages. Two
nebular geometries are considered (thin shell and filled sphere).  We
thus have five parameters in our models. Even though we are interested
in measuring only two of them, O/H and N/O, we still ought to have good
constraints for the other three `uninteresting' parameters: $U$, the
hardness of the ionizing radiation field, and the density
structure. That is why we have chosen a set of emission lines carefully
tailored to constrain those five parameters all at once.

In the following we list these emission lines and explain the reason
behind each of them.  Note that the (reddening-corrected) intensities of
\emph{all} of those lines with respect to \Hb\ are needed to run
\bond.

\begin{figure}
\centering
\includegraphics[width=0.33\columnwidth, trim=10 10 10 7]{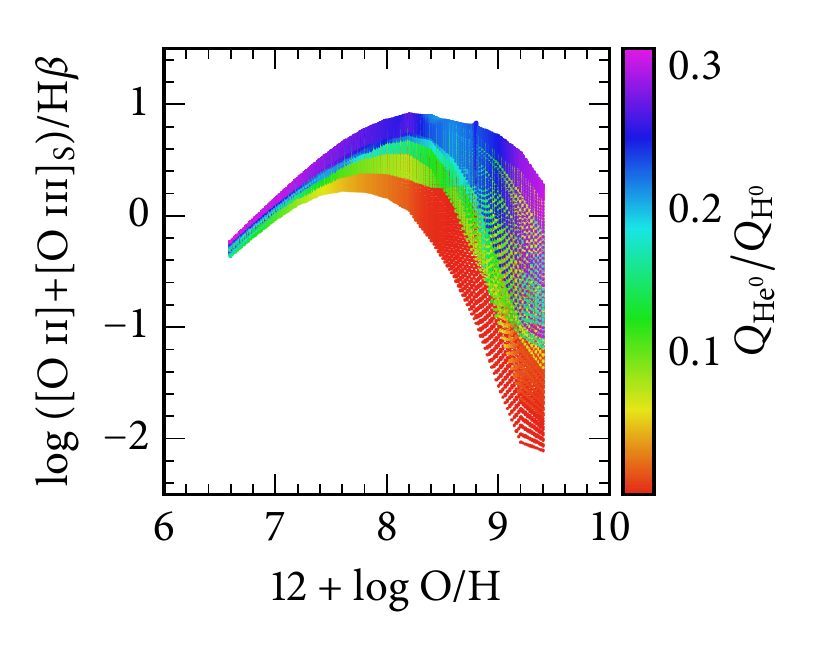}
\includegraphics[width=0.27\columnwidth, trim= 0 10 10 9]{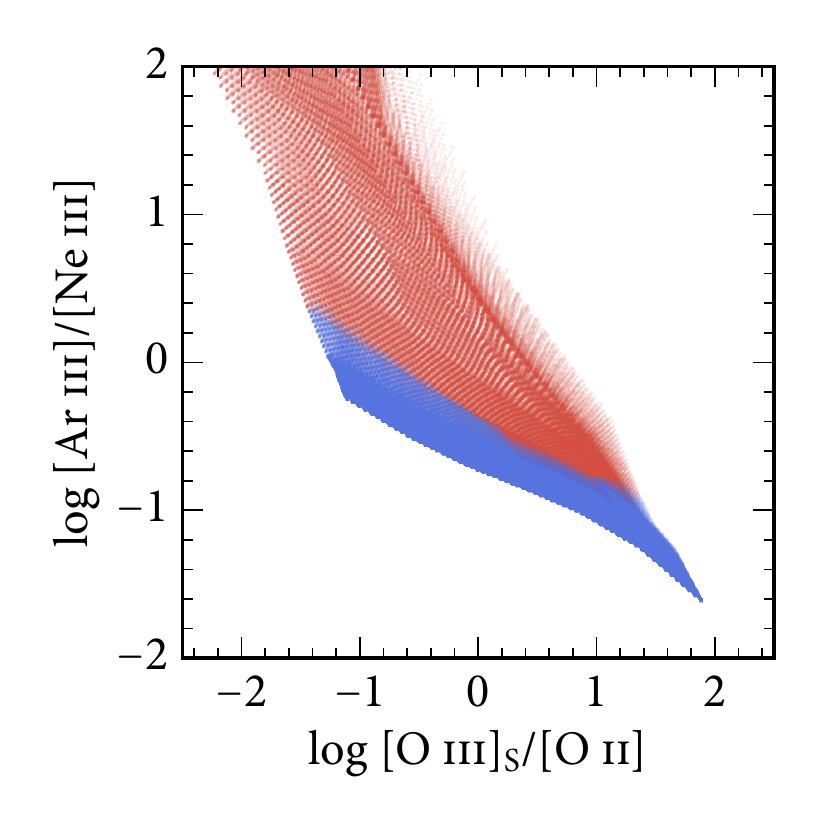}
\includegraphics[width=0.37\columnwidth, trim= 0 10  8 8]{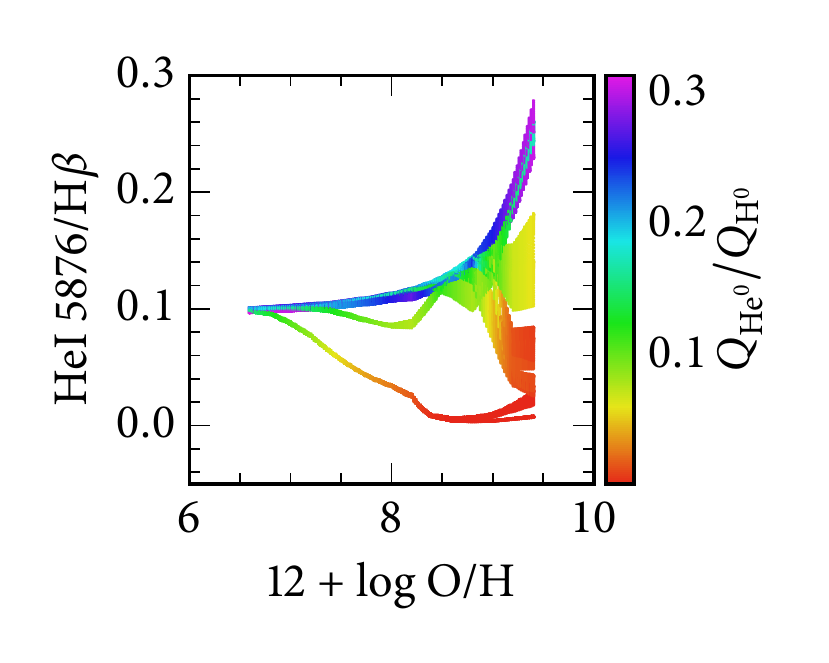}
\caption{\textbf{Left:} $(\oiiis+\oii)/\Hb$ versus O/H coloured by
  \qhe/\qh, which traces the hardness of the ionizing radiation
  field. This represents the two secondary effects considered in
  \bond. First, for a given \qhe/\qh, $(\oiiis+\oii)/\Hb$ maps into two
  different O/H values. We find the correct metallicity branch by using
  the \ariii/\neiii ratio (centre). Second, for high metallicities
  $(\oiiis+\oii)/\Hb$ span almost a decade in O/H. We use \Hei/\Hb
  (right) to find the correct hardness of the ionizing radiation field.
  \textbf{Centre:} \ariii/\neiii versus \oiii/\oii, blue for models in
  the lower metallicity and red for models in the upper metallicity
  branch.  \textbf{Right:} \Hei/\Hb versus O/H coloured by
  \qhe/\qh. \Hei/\Hb can be used as a proxy for \qhe/\qh, except for the
  highest values of \qhe/\qh.  All figures are based on our grid of
  photoionization models and taken from Vale Asari \etal (2016).}
\label{fig:bond-grid}
\end{figure}

\begin{itemize}

\item The strong lines \Hb, \Oii, \Oiii, and \Nii. They were chosen
  because, to first order, the strong line ratios $(\oiii+\oii)/\Hb$,
  $\nii/\oii$, and $\oiii/\oii$ map into O/H, N/O and $U$,
  respectively. We fit line intensities with respect to \Hb, and not the
  latter strong line ratios directly, because we assume that the
  intensities with respect to \Hb follow a Gaussian distribution when we
  calculate likelihood probabilities.
  
\item The semistrong lines \Ariii and \Neiii (plus upper limits for
  auroral lines of \Opp and \Np). The ratio $(\oiii+\oii)/\Hb$ is
  bi-valued with respect to the oxygen abundance (see the inverted U
  shape of the relation in Fig.~\ref{fig:bond-grid}, left). That is why
  some methods impose a fixed relation between N/O versus O/H in their
  photoionization models (e.g.\ Dopita \etal 2013; Blanc \etal
  2015). Since we are interested in inferring both N/O and O/H, we did
  not want to use nitrogen lines to break the bimodality in the oxygen
  abundance.  Restricting our search to emission lines easy to measure
  in typical optical spectra (though not always reported in the
  literature), we found an ideal candidate in the emission line ratio
  \ariii/\neiii. \Arpp and \Nepp are formed in roughly the same zone,
  but the excitation potentials of \ariii and \neiii are very different
  (1.7 and 3.2 eV, respectively), so the ratio of these lines is
  sensitive to the electronic temperature. Argon and neon are primary
  elements and their global abundance ratio is expected to be
  constant. Besides, they are both inert, thus do not suffer dust
  depletion, so their abundance ratio in the ionized gas phase remains
  constant.  Fig.~\ref{fig:bond-grid} (centre) shows the line ratio
  \ariii/\neiii in our grid as a function of \oiii/\oii (the latter
  traces the ionization parameter $U$). The points are colour-coded as
  falling in the lower or upper metallicity branch (blue and red,
  respectively), showing that \ariii/\neiii can break the bimodality in
  the oxygen abundance. An extra help on finding the right metallicity
  branch can come from \emph{upper limits} in auroral lines.

\item The semistrong line \Hei.  This is the crucial part of
  \bond. Since we consider different ionization scenarios, e.g.\
  different spectral energy distributions (SED) of the ionizing
  radiation, we need to infer which of those scenarios is more
  appropriate.  We expect the SED to be different in different \hii
  regions due to ageing (so the most massive stars have disappeared) or
  due to stochastic effects in low luminosity \hii regions when the
  upper part in the stellar initial mass function of the ionizing
  cluster is not fully sampled.  Fig.~\ref{fig:bond-grid} (left) shows
  how the emissivity of $(\oiii+\oii)/\Hb$ depends not only on the
  oxygen abundance (abscissa), but also on he hardness of the ionizing
  radiation field (colour code). Note that at high metallicities a
  single value of $(\oiii+\oii)/\Hb$ can span 1 dex in oxygen abundance
  for different ionizing radiation fields.  Fig.~\ref{fig:bond-grid}
  (right) shows that the \Hei/\Hb line can be used as a proxy of the
  hardness of the ionizing radiation field.

\end{itemize}

\section{Why go Bayesian}
\label{sec:bayes}

First, let us emphasise that the Bayesian inference is \emph{not} what
sets \bond apart. The heart and soul of \bond is the set of carefully
selected emission line ratios, tailor-made both to infer the oxygen and
nitrogen abundances in \hii regions, and to take into account important
secondary parameters. As reasoned in the previous section, this allows
us to (a) break the bimodality in oxygen abundance without any a priori
relation between N/O versus O/H, and (b) consider the (previously
neglected) role of the ionizing radiation field in abundance
determinations.

That said, we have opted for Bayesian inference for good reasons.
Before diving into the most important Bayesian aspects, let us examine
the method's acronym. \bond stands for Bayesian Oxygen and Nitrogen
abundance Determinations. Our method is not called Oxygen and Nitrogen
abundance, Ionization parameter, and Hardness of the ionizing radiation
field Determinations ({\scshape onihd}). The reason why the letters
{\scshape ih} are not in the method's name is the same reason why we
have decided to introduce the {\scshape b} for Bayesian in its name: the
Ionization parameter and Hardness of the ionizing radiation field are
\emph{nuisance} parameters, and the only way to get rid of them
respecting dimensional analysis is by going Bayesian.

Let us lay out the problem to see how its resolution points to Bayesian
inference.  We start with a carefully designed photoionization grid,
finely spaced and spanning a wide range in O/H, N/O, and $U$. It also
considers a few values of hardness of the ionizing radiation field
mimicked by different SED ages. Even though the latter two parameters
($U$ and ionizing radiation field) are important and need to be well
modelled, they are of \emph{secondary} interest. This is why our acronym
shifts from {\scshape onihd} to {\scshape ond}.

How do we get rid of parameters for which we do not care (a.k.a.\
\emph{nuisance} parameters)? If we want to consider the probabilities of
all models in our grid at the same time, then we can simply
\emph{marginalise} over the nuisance parameters. Marginalising is
nothing more than integrating over a parameter. For instance, for a
fixed O/H and N/O, we marginalise over $U$ simply by adding up all the
probabilities of all models of a given O/H and N/O for all $U$.  The
trick is that to integrate, say, in $dU$, the probability density
function (PDF) in the integrand must have physical units of
$U^{-1}$. Ordinary likelihood PDFs (e.g.\ $e^{-0.5 \chi^2}$ for Gaussian
distributions\footnote{
  $\chi^2 = \sum_j {(c_{j} - o_j)^2}/{\sigma_j^2}$, where $o_j$ and
  $\sigma_j$ are the observed line intensity and its uncertainty, and
  $c_{j}$ is the computed line intensity.}) have units which are the
inverse of the observational data being fitted.  So, from a
\emph{pragmatic} point of view, we are obliged to write out the
posterior PDFs, which have the correct physical units when we integrate
over a model parameter.  For a thorough argument on the dimensional
analysis of PDFs, see Hogg (2012), especially the discussion around his
equation 3.

In other words, going Bayesian gives us licence to kill the nuisance
parameters, so we need to add {\scshape b} to {\scshape ond}.
To write the posterior PDFs, we need to spell out our priors.  This has a
two-fold benefit. 
First, we can plug in an informative prior: if we have
empirical evidence that some models are more probable in nature than
others, we can give them more weight by setting the prior probabilities
just right. In our code so far, we have taken the most conservative
approach we can and, following Blanc \etal (2015), we use an
uninformative prior (specifically, a Jeffrey's prior that is
logarithmic in O/H, N/O and $U$).

The second benefit of setting a prior is that we have an explicit
prescription for making a finer grid. The problem of comparing data to a
uniformly spaced grid of models by using a $\chi^2$ likelihood is that
most models will be very distant from the observed data as measured by
the uncertainties $\sigma_j$. If a grid is very very rough, the closest
model might even be a few $\sigma_j$ away from the model with the
highest likelihood. An ad hoc prescription to deal with this problem is
by setting up cooking factors to increase the observed uncertainties,
thus decreasing the distances between observed and computed emission
lines. A cooking factor has no real justification and needs to be
tailored to work for each new data point.  Since we are using a Bayesian
prescription, we can do much better than relying on ad hoc
prescriptions. We simply interpolate our grid where we need---and
interpolation in $\log$ N/O, O/H and $U$ is reasonable once the grid is
fine enough.  The interpolation is informed by posterior probability of
each element in the grid: if an element has a high probability, it is
worth creating more grid points inside its volume\footnote{The algorithm
  to do importance sampling in \bond is the octree sampling, which is
  computationally inexpensive. The reader might be more familiar with
  MCMC samplers, which are more adequate when one has to compute models
  on the fly and does not have a pre-defined grid. In our case, it is
  much more sensible to compute many photoionization models a priori and
  interpolate them on the fly than generating photoionization models on
  the fly. }.

Just a final note on the Bayesian parlance. The outcome of a Bayesian
inference is the posterior PDF for a set of model parameters (say, O/H,
N/O, $U$, hardness of the ionizing radiation field). Since we are
interested only in O/H and N/O, we can integrate out all other
parameters and obtain the \emph{joint} O/H and N/O posterior PDF. The
joint PDF is simply a two-dimensional function that gives the
probability of each point in the O/H versus N/O plane, which is the
ultimate goal of \bond.

However, sometimes it is unpractical to work with the full joint PDF, so
we need a summarised description in the form
$12 + \log \mathrm{O/H} = 8.35 \pm 0.02$. There are many ways to
transform a two-dimensional function into a nominal value and a
dispersion.  One way is to set the nominal value for N/O and O/H to be
the point where the joint PDF is the highest. We call this number the
maximum a posteriori (MAP), because it is calculated after (i.e.\ a
posteriori) the marginalisation of nuisance parameters. For the
dispersion, we can define ellipses of credible regions that encompass,
say, 5, 50, 68 or 95\% of the total joint PDF.

If we want to marginalise away either O/H or N/O (i.e.\ if we are
interested in one of those parameters alone), we can integrate over the
other parameter. Summarising the fully marginalised PDF also opens up a
menu of choices. The nominal value can be taken as the mean, median or
mode (that is, its peak) of the PDF. For the dispersion, the usual
choices are either the 50, 68, or 95 percent equal-tailed or highest
density intervals.  Note that the equal-tailed intervals are related to
the median. The median is the point in the PDF curve where 50 percent of
the probability is to the left and 50 to the right. The equal-tailed
68\% interval is the region of a curve where 16 percent of the
probability is to the right and 16 percent to the left. The highest
density intervals, on the other hand, are related to the mode. The mode
is the point in the curve of highest probability. The 68\% highest
density interval is the region around the mode that adds up to 68
percent of the total probability.  A nice visualisation tool for those
descriptions can be found at
\url{http://www.sumsar.net/blog/2014/10/probable-points-and-credible-intervals-part-one/}.

Another minor nuisance parameter we marginalise away are the
uncertainties. For \ariii/\Hb, \neiii/\Hb and \hei/\Hb, we consider an
extra noise source added in quadrature to the observational
uncertainties which we allow to vary from 2 to 100 percent  of the
line intensity. The extra noise source is needed because, in nature, the
Ar/O and Ne/O ratios may differ somewhat from the ones assumed in our
model grid. Regarding \hei/\Hb, the problem is that our grid is only
coarsely meshed as regards the hardness of the ionizing radiation field.
We then calculate the marginalised likelihood PDF for those lines by
using all values of this extra noise and then marginalising it away. We
do so because we do not expect those lines to be completely correct in
our photoionization models, and they are used only to infer secondary
parameters.

\section{Conclusions}
\label{sec:conclusions}

We have highlighted the main characteristics of \bond, a method based on
a grid of photoionization models to measure oxygen and nitrogen
abundances in giant \hii regions using strong and semistrong lines.  We
show why it is important to consider secondary parameters in abundance
determinations, especially the hardness of the ionizing radiation
field. The SEDs in \hii regions can vary from region to region due to
stellar ageing or to the stochastically sampling of the IMF.  We also
show how one can break the metallicity bimodality without recourse
either to auroral lines or a fixed relation between N/O and O/H. We use
a selective set of emission lines to infer all those
parameters. Finally, we argue why using Bayesian inference is the
correct way (motivated by the dimensional analysis of probability
density functions) to treat the secondary parameters in abundance
determinations.

\section{Acknowledgements}
\label{sec:acknowledgements}

NVA is grateful to Oli Dors for having organised such a useful
workshop. NVA acknowledges the support from Programa de
P\'os-Gradua\c{c}\~ao em F\'{i}sica da UFSC and CAPES/PROAP to attend the
workshop. GS and NVA acknowledge the support from the CAPES CsF--PVE
project 88881.068116/2014-01. The grid of models has been run on
computers from the CONACyT/CB2010:153985, UNAM-PAPIIT-IN107215 and UNAM
Posgrado de Astrof\'{i}sica projects.

\end{document}